\documentclass[aps,prl,reprint,showkeys]{revtex4-1}
\usepackage[utf8]{inputenc}
\usepackage{amsmath}
\usepackage{amsfonts}
\usepackage{amssymb}
\usepackage{hyperref}
\usepackage{simplewick}
\usepackage{color}

\begin{document}
\author{Christian Copetti}
\author{Jorge Fern\'andez-Pend\'as}

\affiliation{Instituto de F\'isica Te\'orica UAM/CSIC, c/Nicol\'as Cabrera 13-15, Universidad Aut\'onoma de Madrid, Cantoblanco, 28049 Madrid, Spain}

\title{Higher spin vortical Zilches from Kubo formulae}
\date{\today}
\begin{abstract}
We compute thermal one point functions in Maxwell's theory sourced by vorticity for the Zilch and its higher spin extensions via the Kubo formalism. This leads to a generalization of the recent results of \citep{Chernodub:2018era} to any spin and their value suggests a relation with possible anomalies for the higher spin tower of currents. 
\end{abstract}

\maketitle

\section{Introduction}
In the last ten years a great deal of effort has been devoted to the elucidation of how nondissipative transport can arise in systems with 't Hooft anomalies. Notably, in the case of fermions, a series of works \cite{Landsteiner:2011iq} established the presence of thermal vortical conductivities whenever the underlying theory has a mixed gravitational anomaly in its chiral currents. While zero temperature effects can be directly related to anomalies by demanding consistence of the hydrodynamic expansion \cite{Son:2009tf}, the thermal case needs the theory to be put on a nontrivial curved manifold \cite{Jensen:2012kj,Stone:2018zel}.

This reasoning seems to be general, and to apply also when the underlying theory has no fermionic excitations. This has led various authors \cite{yamamoto2017photonic,avkhadiev2017chiral} to suppose that the old result of \cite{dolgov1989photonic}, recently revisited by \cite{agullo2017electromagnetic}, should imply a non vanishing vortical conductivity for the helicity current $H_\mu$.
\footnote{
	The helicity current in these works is sometimes defined through $\partial_\mu H^\mu= F_{\mu\nu} \tilde{F}^{\mu\nu} \, , $ which gives the local expression of a Chern-Simons current $H^\mu= 2\epsilon^{\mu\nu\rho\sigma} A_\nu F_{\rho\sigma}$. It is also useful to define a classically conserved helicity current $\tilde{H}^{\mu}=2\epsilon^{\mu\nu\rho\sigma}\left(A_\nu F_{\rho\sigma} - C_\nu \tilde{F}_{\rho\sigma}  \right)$ where the dual gauge field $C$ is defined through $\tilde{F}=d C$. This is the current considered, for example, in \cite{Chernodub:2018era} and the one to bring a close analogue to the axial current for fermions.
}

This current is however not gauge invariant and contrasting results exist in the literature regarding its thermal expectation value \cite{yamamoto2017photonic,avkhadiev2017chiral,Chernodub:2018era}. However, higher spin gauge invariant currents can be built from the helicity current and one expects, as it happens in two dimensional theories, that these should also show quantum correction to their conservation laws.

Following the original nomenclature we will call them higher spin Zilches $Z^{(s)}_{\mu_1...\mu_s}$.
One would then expect these currents to develop thermal nondissipative transport in the presence of vorticity, through the presence of a term
$\sigma^{(s)}_Z u_{(\mu_1} ... u_{\mu_{s-1}} \omega_{\mu_s)}$ 
in its hydrodynamic expansion, with $\sigma_Z^{(s)} \sim T^{s+1}$ by dimensional analysis.

The existence of such terms has already been shown for the spin-3 Zilch by \cite{Chernodub:2018era} through the quantization of Maxwell's theory in a rotating cylinder, for which they have found
\begin{equation}
\langle Z_{00i}\rangle_{\beta,\Omega}= \frac{8}{45} \pi^2 T^4 \Omega_i \, .
\end{equation}

The aim of this work is to generalize these result to the whole tower of Zilches through the Kubo formalism. This extension may be interesting, since it was already pointed out that the spin three current is potentially measurable in an experimental setup \cite{tang2010optical}.

We carry out the computations, showing that indeed such vortical responses are present for all odd spins, being equal to
\begin{equation}
\sigma_Z^{(s)}= \frac{4}{\pi^2} \left(2\pi T \right)^{s+1} \frac{\lvert B_{s+1} \rvert}{s+1} \, ,
\end{equation}
where $B_s$ are the Bernoulli numbers.
This confirms the results of \cite{Chernodub:2018era} through a different method and opens up a new interesting avenue for the field on anomalous transport i.e. to relate higher spin 't Hooft anomalies to the transport properties of these higher spin currents.

In this respect one should notice that our result closely resembles the one obtained in \cite{Bonora:2008nk,Bonora:2008he}. The authors compute the expectation value of the 1+1 dimensional free bosonic currents as part of an old program aiming to relate the moments of Hawking radiation with the gravitational anomalies of an effective description of the horizon physics \cite{robinson2005relationship,iso2007higher,iso2007fluxes}.

It is known that such currents give a representation of the $W_{\infty}$ algebra, as shown by \cite{bakas1990bosonic}. In that case, however, the cohomological problem for the consistent $W_\infty$ currents has only trivial solutions for $s>2$. Thus the possible anomalies in the higher spin currents can be adsorbed into their definition by a proper choice of a local counterterm.

In the two dimensional case, then, the nonvanishing one-point function can be attributed to the nonhomogeneous transformation properties of the higher spin currents under general conformal transformations (they are only quasi-primary operators in the CFT). The final effect is very similar to the Schwartzian transformation of the stress tensor, which gives the starting point for the derivation of the trace and diffoemorphism anomalies in chiral CFTs, but the mechanism for higher spin is different since the identity operator cannot appear in the OPE of the higher spin currents with the stress tensor. The anomalous transformation law is, however, a purely quantum effect given by the normal ordering of the current operators and, in this sense, it could still make up interesting and robust observables in a variety of physical systems.

In the four dimensional case the mechanism could be similar, and it could maybe be proven by performing an appropriate reduction to two dimensions. We provide some details on how it could be done in the conclusions.

The letter is organized as follows: in the first section, we review the spin-3 computation for the Zilch; in the second section, we show how to define odd spin Zilches and compute their vortical conductivities in the Kubo formalism, and we conclude with few remarks on future directions for investigation. In order to make the letter more readable, long computations are relegated to the supplemental material. Regarding notation, we mostly work in Euclidean signature, dividing the 4-momentum as $k_\mu=(\omega_n,\vec{k})$ where $\omega_n= 2\pi T n$ are bosonic Matsubara frequencies. In much the same way we split the spacetime indexes in spatial components, denoted by Latin letters $i,j,k...$ and the time direction denoted by $0$.

\section{Vortical Zilch}
As is the case for free theories, Maxwell theory itself presents towers of higher spin currents, which are built starting from the stress tensor
\begin{equation}
T_{\mu\nu}= F_{\mu\alpha} {F_{\nu}}^\alpha - \frac{1}{4}g_{\mu\nu} F^2 \, ,
\end{equation} 
and the spin three Zilch
\begin{equation}
Z_{\mu\nu\rho}= {F_{(\mu}}^\alpha \overset{\leftrightarrow}{\partial_\rho} \tilde{F}_{\nu) \alpha} \, , \label{Zilchdefo}
\end{equation}
where $\overset{\leftrightarrow}{\partial_\rho} = \frac{1}{2}\left( \overset{\rightarrow}{\partial_\rho} - \overset{\leftarrow}{\partial_\rho} \right)$ and $\tilde{F}_{\mu\nu}=\epsilon_{\mu\nu\rho\sigma}F^{\rho\sigma}$ is the dual field strength. From now on, in order to avoid cluttering,  complete symmetrization in all of the Zilches indexes is implicit
\footnote{
		Our convention for symmetrization is
		\begin{equation}
		V_{(\mu_1...\mu_s)}=\frac{1}{s!}\sum_{\pi \in S_n} V_{\pi(\mu_1)...\pi(\mu_s)} \, .
		\end{equation} 
}. 
The Zilch fulfills
\begin{equation}
\partial^\mu Z_{\mu\nu\rho}= {Z^\mu}_{\mu\rho} =0  \, .
\end{equation} 
This current was analyzed in various early works \cite{lipkin1964existence, kibble1965conservation}, ultimately being deemed to be unimportant. Notice that both the Zilch current and the stress tensor (as well as their higher spin counterparts we will define) are invariant under the duality symmetry of Maxwell's theory $ F \to \tilde{F}, \ \ \tilde{F} \to - F$.

A renewed interest was sparked by \cite{Chernodub:2018era}, where it was shown to possess a nontrivial one point function at finite temperature in the presence of vorticity. In this section, we will reproduce the results of \cite{Chernodub:2018era} by using the Kubo formalism. We wish to compute the thermal response to vorticity $\sigma_Z$, defined through
\begin{equation}
Z_{\mu\nu\rho} = \sigma_Z u_{(\mu} u_{\nu} \omega_{\rho)} + \dots \, ,
\end{equation}
or, in the fluid rest frame,
\begin{equation}
Z_{00i}= \frac{1}{3}\sigma_Z \Omega_i \, . \label{Zonethirdsigma}
\end{equation}

At this point it is worth noting that our definition of Zilch \eqref{Zilchdefo} does not coincide with the original definition of Lipkin \cite{lipkin1964existence} which is used in \cite{Chernodub:2018era}, but it differs from that one by the curl of the Poynting vector. One can show using the results of \cite{Chernodub:2018era} that such term gives no contribution to the vortical conductivity. Another argument, perhaps less stringent, is to notice that, in a rotating ensemble, the only preferred direction is given by the vorticity vector. So, to linear order, $\langle T_{0i} \rangle_{\beta, \Omega} \sim c(T) \Omega_i $ and thus its curl cannot give contributions proportional to the vorticity itself.
       
Once this has been clarified we can use linear response theory to compute the relevant transport coefficients. In this setup the vortical conductivity is given by the Kubo formula
\begin{equation}
\sigma_Z= 6 \lim_{\vec{p}\to 0} \frac{-i}{2 p_k} \epsilon^{ijk} \bigl( G_{00i,0j}(p) + C_{00i,0j}(p)\bigr) \, ,
\end{equation}
where 
\begin{align}
G_{\mu\nu\rho,\alpha\beta}(x-y) &= -i \big\langle [ Z_{\mu\nu\rho}(x), T_{\alpha\beta}(y)] \big\rangle \Theta(t-t') \, ,\label{2pf} \\
C_{\mu\nu\rho,\alpha\beta}(x-y)&= 2 i \left\langle \frac{\delta Z_{\mu\nu\rho}(x) }{\delta g_{\alpha \beta}(y)} \right\rangle \label{seagull}  \, ,
\end{align}
are the retarded Green's function and the seagull term
\footnote{
The seagull contribution only comes from the covariantization of derivatives in the Zilch. Other contributions, which are due to the metric dependence of $\tilde{F}$ and the index contractions give terms independent of the external momentum.
}, respectively, and the factor of $6$ comes from \eqref{Zonethirdsigma} and expressing the response in terms of the gravitomagnetic field instead of the vorticity. The factor in \eqref{seagull} comes from lowering the stress tensor with $2i\frac{\delta}{\delta g}$. From now on, we will switch to momentum space where we can use the standard relation
\begin{equation}
G_{\mu\nu\rho,\alpha\beta}(0,\vec{p})= - G^E_{\mu\nu\rho,\alpha\beta}(0,\vec{p}) \, ,
\end{equation}
where $G^E$ is the Euclidean Green's function and a similar relation holds for the seagull term. 
 The calculation is greatly simplified by using Wick contractions of the field strength $F_{\mu\nu}$ in order to write the momentum space integrals
\begin{align}
\contraction{F}{_{\mu\nu}(p)}{}{F} F_{\mu\nu}(p) F_{\rho\sigma}(q) &= (2\pi)^{-4} \delta(p+q) L_{\mu\nu\rho\sigma}(p)\, , \\
 L_{\mu\nu\rho\sigma}(q) &= -\frac{4}{q^2} q_{[\mu}g_{\nu][\sigma}q_{\rho]} \, .
\end{align}
Using this, one arrives at the Feynman integrals
\begin{widetext}
\begin{align}
G_{\mu\nu\rho,\alpha\beta}(p)&= \frac{-i}{2 \beta}\sum_n \int \frac{d^3 \vec{k}}{(2\pi)^3} {\epsilon_{\nu}}^{\sigma \tau \gamma} \left( \frac{p}{2} - k \right)_\rho  \  \left[ L_{\mu\sigma\alpha\xi}(k){L_{\tau\gamma\beta}}^\xi(p-k) + L_{\mu\sigma\beta\xi}(k){L_{\tau\gamma\alpha}}^\xi(p-k) \right] \, , \\
C_{\mu\nu\rho,\alpha \beta}&= i p^\gamma \frac{1}{\beta}\sum_n \int \frac{d^3 \vec{k}}{(2\pi)^3}\left[{\epsilon_{\nu(\alpha}}^{\sigma \tau}\delta_{\beta)\rho} L_{\mu\gamma\sigma\tau}(k) -{\epsilon_{\nu\gamma}}^{\delta\tau}\delta_{\rho(\alpha} L_{\mu\beta)\delta\tau} (k)\right] \, ,
\end{align}
\end{widetext}
where we use Lorentz indexes for simplicity and $k_\mu=(\omega_n,\vec{k})$ and we drop terms proportional to $g_{\alpha \beta}$ from the equations since we are only interested in the case $\alpha \beta = 0 j$.

Once the correct expressions in momentum space are inserted and the linear order in momentum $p$ is extracted, the final result can be written in terms of the following family of divergent integrals
\begin{equation}
I_D^{(a,b,c)} = \frac{1}{\beta}\sum_n\int \frac{d^D k}{(2\pi)^D} \frac{ \lvert \vec{k} \rvert^{2a} \omega_n^{2c}}{(\omega_n^2 + \lvert \vec{k} \rvert^2)^b} \, ,
\end{equation}
where $\omega_n= 2 \pi n T$ are the bosonic Matsubara frequencies. The integral is divergent, but it can be regulated following \cite{Golkar:2012kb, avkhadiev2017chiral} via dimensional regularization, followed by zeta-function regularization of the Matsubara sums. The finite result reads:
\begin{widetext}
\begin{equation}
I_D^{(a,b,c)}= T^{D +1 +2(a-b+c)}2^{-D/2 +1}(2 \pi)^{D/2 +2(a-b+c)} \frac{\Gamma(a + D/2)\Gamma(b-a-D/2)}{\Gamma(D/2) \Gamma(b)}\zeta(-D -2(a-b+c))\, .
\end{equation}
\end{widetext}
In terms of these integrals, the contributions to the vortical Zilch are given by
\begin{align}
G_{00i,0j}(p) =& i \epsilon_{ijk}p^k {\cal I}_3^G \, , \\
C_{00i,0j}(p) =& i \epsilon_{ijk}p^k {\cal I}_3^C \, ,
\end{align}
where
\begin{align}
{\cal I}_3^G &= \frac{2}{3} \left(\frac{1}{3} I_{3}^{(1,1,0)} - 2 I_3^{(1,2,1)} - I_3^{(0,1,1)} + 2 I_{3}^{(0,2,2)} \right) \nonumber \\
&= \frac{8}{135} \pi^2 T^4 \, , \\
{\cal I}_3^C &= -\frac{2}{3}\left( \frac{1}{3}I_3^{(1,1,0)} - I_3^{(0,1,1)} \right) = \frac{4}{135} \pi^2 T^4 \, .
\end{align}
All these integrals are in fact proportional to each other, and summing them up gives
\begin{equation}
\sigma_Z = \frac{8}{15} \pi^2 T^4 \, ,
\end{equation}
reproducing the result of \cite{Chernodub:2018era} once we convert to their conventions 
\footnote{
	The changes are as follows: the current given in \cite{Chernodub:2018era} coincides with our definition of $Z_{00i}$ after explicit use of the Maxwell's equations, while our vortical conductivity is three times larger than theirs, due to having defined it in a symmetric and covariant way.
}.

\section{Generalization to higher spins}

At this point we would like to generalize the result of the previous section to the whole tower of higher spin currents constructed from the Zilch. We would expect, on general grounds, such one point function to display contributions going like $T^{s+1}$ in a thermal background in the presence of vorticity. A related but different approach to compute the response of higher spin gauge fields was developed in \cite{Huang:2018aly} using the chiral kinetic theory of \cite{Stephanov:2012ki}.

For starters, we can already guess that none of the even spin currents would give rise to nonvanishing vortical Zilches. In fact, a moment of thought shows us that they should come with Matsubara sums of odd powers of the frequencies $\omega_n$, which vanish identically in our regularization scheme.

Restricting the analysis only to odd spins $s= 3 + 2n$, we can proceed with the construction of the higher spin Zilches in position space, whose explicit form is needed for the computation of the contact terms. One can guess the answer to be
\footnote{
	As a matter of fact, this is not the only possible choice for the higher spin currents, since one can mix them with lower spin ones and change their normalization factors. A more complete definition would involve explicit examination of their quantum algebra, which we leave for future work.
}
\begin{equation}
Z^{(s)}_{\mu_1...\mu_s} = {F_{(\mu_1}}^\alpha \overset{\leftrightarrow}{\partial}_{\mu_2} ... \overset{\leftrightarrow}{\partial}_{\mu_{s-1}} \tilde{F}_{\mu_s) \alpha} \, .\label{higherzilches} 
\end{equation}
The Zilches are indeed conserved and traceless on-shell
\begin{align}
\partial^\mu Z^{(s)}_{\mu \mu_2 ... \mu_s} &= 0 \, , \\
{Z^{(s)}}^\mu_{\mu \mu_3 ... \mu_s} &= 0 \, .
\end{align}
The proof of these properties can be found in the supplemental material.

 The aim of this section is to see whether, in the presence of vorticity, the following linear response expansion holds:
\begin{equation}
Z^{(s)}_{\mu_1...\mu_s}= \sigma^{(s)}_Z u_{(\mu_1} ... u_{\mu_{s-1}} \omega_{\mu_s)} + \dots \, .
\end{equation} 
This leads in the rest frame to the following Kubo-type relation for the higher spin vortical conductivity
\begin{equation}
\sigma^{(s)}_Z=2s \lim_{\vec{p}\to 0} \frac{i}{2 p_k} \epsilon^{ijk} \left( G_{00...0i,0j}(p) + C_{00...0i,0j}(p)\right) \, ,
\end{equation}
where the two point functions and the seagull term are the obvious generalizations of \eqref{2pf} and \eqref{seagull} and the $2s$ factor comes by correctly taking into account symmetrization and the relation between vorticity and gravitomagnetic field. The only minor technical difficulty in generalizing the computation, once the explicit form of the higher spin Zilches is given, is the computation of the seagull term.

In order to do this, one should rewrite the flat space expression with covariant derivatives and take functional derivatives with respect to the metric.  The position space expression for the contact term is rather complicated and is reported in the supplementary material. In practice, it suffices to say that most terms are made up of a field strength and a dual contracted by one index. However, in momentum space, all these terms give a vanishing contribution to the seagull diagram due to the identity
\begin{equation}
\epsilon^{\nu\alpha\beta\gamma}L_{\mu\alpha\beta\gamma}(q)=0 \, .
\end{equation}
Finally there is one last contribution where such terms are not contracted, which can be recast into the compact expression
\begin{widetext}
\begin{equation}
\begin{aligned}
C_{\mu_1... \mu_s, \alpha \beta}(x-y)= -\frac{(s-2)}{2} \left( F_{(\mu_1\gamma} \overset\leftrightarrow{\partial}_{\mu_2} ... \overset\leftrightarrow{\partial}_{\mu_{s-2}} \tilde{F}_{\mu_{s-1}}{ }^\delta 
 - {F_{(\mu_1}}^\delta \overset\leftrightarrow{\partial}_{\mu_2} ... \overset\leftrightarrow{\partial}_{\mu_{s-2}} \tilde{F}_{\mu_{s-1}\gamma}  \right) \frac{\delta \Gamma^\gamma_{\mu_s)\delta}}{\delta g^{\alpha \beta}(y)} + {\cal O} ( \partial \Gamma ) \, .
\end{aligned}
\end{equation}
We can then read the contributions to the higher vortical conductivities from
\begin{align}
G_{\mu_1...\mu_s,\alpha \beta}(p) = \frac{-i}{2 \beta}\sum_n \int \frac{d^3 \vec{k}}{(2\pi)^3}(-)^{(s-3)/2} {\epsilon_{\mu_1}}^{\sigma \tau \gamma} \left(\frac{p}{2}-k\right)_{\mu_2}... \left(\frac{p}{2}-k\right)_{\mu_{s-1}} \bigg[& L_{\mu_s\sigma\alpha\xi}(k){L_{\tau\gamma\beta}}^\xi(p-k) \nonumber \\
&+ L_{\mu_s\sigma\beta\xi}(k){L_{\tau\gamma\alpha}}^\xi(p-k) \bigg] \, ,
\end{align}
\begin{equation}
C_{\mu_1... \mu_s, \alpha \beta}(p)= (s-2) i p^\gamma \frac{1}{\beta}\sum_n \int \frac{d^3 \vec{k}}{(2\pi)^3}(-)^{(s-3)/2}k_{\mu_2}... k_{\mu_{s-2}}\left[{\epsilon_{\mu_1(\alpha}}^{\sigma \tau}\delta_{\beta)\mu_{s-1}}L_{\mu_s\gamma\sigma\tau}(k) -{\epsilon_{\mu_1\gamma}}^{\delta\tau}\delta_{\mu_{s-1}(\alpha} L_{\mu_s\beta)\delta\tau} (k)\right] \, .
\end{equation}
\end{widetext}
Setting $(\mu_1,...,\mu_s)=(0,0...,i)$ has essentially the effect of multiplying the contributions from the $s=3$ case by an appropriate power of Matsubara frequencies. The thermal contributions then read
\begin{align}
G_{00...0i,0j}(p)&= i \epsilon_{ijk}p^k {\cal I}_s^G \, , \\
C_{00...0i,0j}(p)&= i \epsilon_{ijk}p^k {\cal I}_s^C \, ,
\end{align}
where, setting $s=3 + 2n$ and working with odd spins only, we obtain
\begin{align}
{\cal I}_s^G = \frac{2}{s} \bigg[&\frac{s-2}{3} \left(I_{3}^{(1,1,n)} - 2 I_3^{(1,2,n+1)}\right) \nonumber \\
&- \frac{4}{3} I_3^{(1,2,n+1)} - I_3^{(0,1,n+1)} + 2 I_{3}^{(0,2,n+2)} \bigg] \, , \\
{\cal I}_s^C = -\frac{2}{s} (& s - 2 ) \left( \frac{1}{3} I_3^{(1,1,n)} - I_3^{(0,1,n+1)} \right) \, .
\end{align} 
The factors of $s-2$ come from integrals in which the spatial index is on one of the momenta, since there are $s-2$ ways in which this can happen, while the factor of $1/s$ is a remnant of the symmetrization procedure.
After regularization the integrals combine to give
\begin{equation}
\sigma_Z^{(s)}= \frac{4}{\pi^2} \left(2\pi T \right)^{s+1}(-)^{(s-1)/2} \frac{B_{s+1}}{s+1} \, , \label{higherspinZilch}
\end{equation}
where $B_n$ are the Bernoulli numbers. We should note that, since even Bernoulli numbers are alternating in sign, the result is positive for every odd spin greater than one. This formula is valid for odd spins $s= 3 + 2n$, while for even spin we expect the result to vanish due to the zeta function regularization of the Matsubara sums.
As we have pointed out in the introduction, this result closely resembles the one obtained for the one point functions of the $W_{\infty}$ currents of the free complex boson.

\section{Discussion and conclusions}
We have explicitly computed the vortical conductivities for Zilches of spin higher than three in the Kubo formalism. Let us briefly comment on the physical interpretation of our results, especially of equation \eqref{higherspinZilch}.

First, one may take the limit $s \to 1 $ of the expressions for the conductivities $\sigma_Z^{s}$, obtaining $\sigma_Z^{1}=4/3 T^2$. One may desire to interpret this as a regularized computation of the vortical conductivity for the helicity current. In fact, our definition of the Zilches allows to take a smooth $s \to 1$ limit, at least for the charges, and thus analytically continue the expressions to lower spins. However, while with our normalization the definition of the conserved charges
$Q^{s}_Z= \int d^3x Z^{s}_{0..0}$ coincides with the ones in \cite{Chernodub:2018era}, the $s \to 1$ limit of the currents $Z_{0...0i}$ does not coincide with the expression for the helicity current in any simple gauge.

Another important point to be clarified is whether the higher spin responses are actually stemming from anomalies. As we have mentioned in the introduction, some results are available in $1+1$ dimensions, where the $W$ algebras do not have higher spin diffoemorphism anomalies. One can however define $W$ currents which transform covariantly under diffeomorphisms, but have a nontrivial conservation law. In this sense one can recover the expressions for the one point functions (at least for relatively small spin, in which case the conservation law is actually computable) by an argument along the lines of \cite{Jensen:2012kj} or \cite{Stone:2018zel}.

Because of this, it would be very interesting to develope a systematic way to compare the four dimensional result in the presence of vorticity to the two dimensional ones, much in the same way as the case of strong magnetic field, where the low energy physics is essentially given by the $1+1$ dimensional description of the Lowest Landau Level.
One could for example try to study the system on ${\rm I\!R}^2 \times S^2$ with the vorticity playing the role of a chemical potential for the $SO(3)$ symmetry and perform a Kaluza-Klein reduction. Intuitively, the operator spectrum should contain an $SO(2)$ doublet of fields $a_i$ which are scalars from the point of view of the two dimensional theory, and strongly reminiscent of the $W_\infty$ case. We will however leave such reflections for future works.

Furthermore, one needs to recall that, at least in dimensions higher than two, higher spin theories are free \cite{Maldacena:2011jn}. Thus, we expect that the inclusion of interactions, that will softly break the higher spin symmetry, would cause our results to change. In a different context, however \cite{Copetti:2016ewq}, it was shown that similar breaking only have the effect of renormalizing the gauge coupling of the current which appear in the anomaly polynomial.

\section{Acknowledgements}
The authors would like to thank Karl Landsteiner for introducing us to the problem and encouraging the publication of our results. This work is supported by FPA2015-65480-P and by the Spanish Research Agency (Agencia Estatal de Investigación) through the grant IFT Centro de Excelencia Severo Ochoa SEV-2016-0597. The work of J.F.-P. is supported by fellowship SEV-2012-0249-03. The work of C.C. is funded by Fundaci\'on La Caixa under ``La Caixa-Severo Ochoa'' international predoctoral grant.

\bibliography{AnomTrans}{}

%%%%%%%%%% Merge with supplemental materials %%%%%%%%%%
%\pagebreak
\onecolumngrid
\newpage
%\widetext

\begin{center}
\textbf{\large Supplemental Materials: Higher spin vortical Zilches from Kubo formulae}
\end{center}
%%%%%%%%%% Merge with supplemental materials %%%%%%%%%%
%%%%%%%%%% Prefix a "S" to all equations, figures, tables and reset the counter %%%%%%%%%%
\setcounter{equation}{0}
\setcounter{figure}{0}
\setcounter{table}{0}
\setcounter{page}{1}
\makeatletter
\renewcommand{\theequation}{S\arabic{equation}}
\renewcommand{\thefigure}{S\arabic{figure}}
\renewcommand{\bibnumfmt}[1]{[S#1]}
\renewcommand{\citenumfont}[1]{S#1}
%%%%%%%%%% Prefix a "S" to all equations, figures, tables and reset the counter %%%%%%%%%%

\section{Regularization of $I^{(a,b,c)}$}
Here we briefly show how to regulate the various divergent integral appearing in the thermal computations. We will use a mix of $\zeta$-function and dimensional regularization. This is quite suitable, since the spatial integral happens to be in an odd number of dimensions and thus it will get automatically regulated.
We wish to compute
\begin{equation}
I_D^{(a,b,c)} = \frac{1}{\beta}\sum_n\int \frac{d^D k}{(2\pi)^D} \frac{\lvert \vec{k} \rvert^{2a} \omega_n^{2c}}{(\omega_n^2 + \lvert \vec{k} \rvert^2)^b} \, .
\end{equation}
We start with the spatial part, which can be expressed in terms of
\begin{equation}
I_D^{(a,b)}(\Delta) = \int \frac{d^D \vec{k}}{(2\pi)^D} \frac{\lvert \vec{k} \rvert^{2a}}{( \Delta + \lvert \vec{k} \rvert^2 )^b} \, .
\end{equation}
Changing to spherical coordinates and using the integral representation of Euler's beta function
\begin{equation}
B(u,v)=\int_0^\infty dy y^{u-1} (1+y)^{-v-u} \, ,
\end{equation}
gives immediately
\begin{equation}
I^{(a,b)}(\Delta)= \frac{\Delta^{D/2 + a -b}}{(4\pi)^{D/2} \Gamma(D/2)} \frac{\Gamma(a+D/2) \Gamma(b-a-D/2)}{\Gamma(b)} \, .
\end{equation}
Our initial integral has now become
\begin{equation}
I_D^{(a,b,c)} = T (2\pi T)^{D +2(a-b+c)} \left(1 + (-)^{2c} \right) \sum_{n=0}^\infty n^{D +2(a-b+c)} I_D^{(a,b)}(1) \, .
\end{equation}
The final sum is regulated by using zeta function regularization $\zeta(s)=\sum_{n=1}^{\infty} \frac{1}{n^s}$. Notice that the whole expression vanishes when $c$ is half integer, which is the case for even spin Zilches. 
Algebraic simplifications then give
\begin{equation}
I_D^{(a,b,c)} = T^{D +1 +2(a-b+c)}2^{-D/2 +1}(2 \pi)^{D/2 +2(a-b+c)} \frac{\Gamma(a + D/2)\Gamma(b-a-D/2)}{\Gamma(D/2) \Gamma(b)}\zeta(-D -2(a-b+c))\, ,
\end{equation}
which is perfectly well defined for $D=3$.

In order to obtain the final expression in the main text, one uses that
\begin{equation}
\zeta(-s)= (-)^s \frac{B_{s+1}}{s+1} \, ,
\end{equation}
where $B_n$ are the Bernoulli numbers, e.g. $B_2= 1/6, \ \ B_4= -1/30$ etc. Direct computation shows that all the relevant integrals are proportional to each other with proportionality constants independent of the powers $c$ of the frequency.

\section{Construction of the higher Zilches and form of the contact terms}
In this section we explicitly verify the conservation for the higher Zilch currents \eqref{higherzilches} and derive from these currents the contact terms for the Kubo formula computation. We omit for simplicity unimportant normalization factors.
We start by showing that the current is conserved $\partial^\mu Z^{(s)}_{\mu \mu_2...\mu_s}=0$. In order to do this, one expands
\begin{equation}
Z^{(s)}_{\mu \mu_2...\mu_s} = \frac{1}{s}\left({F_{\mu}}^\alpha \overset{\leftrightarrow}{\partial}_{(\mu_2} ... \overset{\leftrightarrow}{\partial}_{\mu_{s-1}} \tilde{F}_{\mu_s) \alpha} + {F_{(\mu_2}}^\alpha \overset{\leftrightarrow}{\partial}_{\mu_3} ... \overset{\leftrightarrow}{\partial}_{\mu_{s})} \tilde{F}_{\mu \alpha}   \right) 
+ \frac{s-2}{s} {F_{(\mu_2}}^\alpha \overset{\leftrightarrow}{\partial}_{\mu} \overset{\leftrightarrow}{\partial}_{\mu_3} ... \overset{\leftrightarrow}{\partial}_{\mu_{s-1}} \tilde{F}_{\mu_s) \alpha} \, .
\end{equation}
When applying the divergence, the last term vanishes due to the equation of motion $\square F_{\mu\nu}=\square \tilde{F}_{\mu\nu}=0$. This happens since its contraction with the two sided derivative reads $ \overset{\leftrightarrow}{\partial}_{\mu}^+ \overset{\leftrightarrow}{\partial}^\mu= \overset{\leftrightarrow}{\square}$ , where $ \overset{\leftrightarrow}{\partial}_{\mu}^+ = \overset{\rightarrow}{\partial_\mu} + \overset{\leftarrow}{\partial_\mu}$.
The first two terms give a contribution
\begin{equation}
\partial^\mu Z^{(s)}_{\mu \mu_2...\mu_s}= \frac{1}{s}\left({F_{\mu}}^\alpha \overset{\leftrightarrow}{\partial}_{(\mu_2} ... \overset{\leftrightarrow}{\partial}_{\mu_{s-1}} \partial^\mu \tilde{F}_{\mu_s) \alpha} + \partial^\mu{F_{(\mu_2}}^\alpha \overset{\leftrightarrow}{\partial}_{\mu_3} ... \overset{\leftrightarrow}{\partial}_{\mu_s)} \tilde{F}_{\mu \alpha}   \right) \, , 
\end{equation} 
where we have already dropped the other combination which vanishes due to Maxwell's equation $\partial_\mu F^{\mu\nu}= \partial_\mu \tilde{F}^{\mu\nu}=0$. 

The remaining terms have to be manipulated a bit in order to show that they cancel. To do this one uses the Bianchi identity and the antisymmetry in $\alpha \mu$ to substitute $ \partial_\mu F_{\mu_2 \alpha}$ by $- \frac{1}{2} \partial_{\mu_2} F_{\mu \alpha}$ and the same for $\tilde{F}$. This results in
\begin{equation}
\partial^\mu Z^{(s)}_{\mu \mu_2...\mu_s}= -\frac{1}{2s} \partial_{(\mu_2} \left( F^{\mu\alpha} \overset{\leftrightarrow}{\partial}_{\mu_3} ... \overset{\leftrightarrow}{\partial}_{\mu_s)} \tilde{F}_{\mu \alpha}\right).
\end{equation}
Since the number of double sided derivatives is odd, the expression is both symmetric and antisymmetric in $F \leftrightarrow \tilde{F}$ so it vanishes.

Tracelessness follows in a similar way. In fact, using the equation of motion we can rewrite the trace of the Zilches as
\begin{equation}
\begin{aligned}
{{Z^{(s)}}^\mu}_{\mu \mu_3 ... \mu_s} &=\frac{2}{s (s-1)} F^{\mu\alpha} \overset{\leftrightarrow}{\partial}_{(\mu_3} ... \overset{\leftrightarrow}{\partial}_{\mu_s)} \tilde{F}_{\mu \alpha} - \frac{(s-2)(s-3)}{4 s (s-1)} \partial_\mu {F_{(\mu_3}}^\alpha \overset{\leftrightarrow}{\partial}_{\mu_4} ... \overset{\leftrightarrow}{\partial}_{\mu_{s-1}} \partial^\mu \tilde{F}_{\mu_s) \alpha} \, \\
&+ \frac{s-2}{s (s-1)} \left({F_{\mu}}^\alpha \overset{\leftrightarrow}{\partial}_{(\mu_3} ... \overset{\leftrightarrow}{\partial}_{\mu_{s-1}} \partial^\mu\tilde{F}_{\mu_s) \alpha} - \partial^\mu{F_{(\mu_3}}^\alpha \overset{\leftrightarrow}{\partial}_{\mu_4} ... \overset{\leftrightarrow}{\partial}_{\mu_s)} \tilde{F}_{\mu \alpha} \right) \, ,
\end{aligned}
\end{equation}
which is immediately seen to vanish term by term once the Bianchi identity is used to simplify the second line. Notice that is is critical for the spin to be odd in order for the computation to work out.
Having constructed a conserved spin-s Zilch in flat spacetime we wish to extend it to the curved case to extrapolate the contact terms relevant to our calculation.

Now we move on to compute the contact term, by making the partial derivatives covariant. We will work only at linear level in the curved metric and, in order to do this, it is expedient to rewrite the currents as
\begin{equation}
Z^{(s)}_{\mu_1...\mu_s}= \sum_{k=0}^{s-2} c_{s,k} \partial_{(\mu_2}...\partial_{\mu_{k}} {F_{\mu_1}}^\alpha \partial_{\mu_{k+1}}...\partial_{\mu_{s-1-k}} \tilde{F}_{\mu_s) \alpha} \, ,
\end{equation}
where $c_{s,k}= \frac{(-)^k}{2^{s-2}} {{s-2}\choose{k}}$. To covariantize we simply replace partial derivatives with covariant ones, and to linear order we only have to worry of a single covariant derivative at a time, so that we may write
\begin{equation}
Z^{(s)}_{\mu_1...\mu_s}= \sum_{k=0}^{s-2} c_{s,k} \sum_{i=0}^k \partial_{(\mu_2}...\nabla_{\mu_i}...\partial_{\mu_{k}} {F_{\mu_1}}^\alpha \partial_{\mu_{k+1}}...\partial_{\mu_{s-1-k}} \tilde{F}_{\mu_s) \alpha} -\partial_{(\mu_{k+1}}...\partial_{\mu_{s-1-k}}  {F_{\mu_1}}^\alpha \partial_{\mu_2}...\nabla_{\mu_i}...\partial_{\mu_{k}} \tilde{F}_{\mu_s) \alpha} \, ,
\end{equation}
where the minus sign is a consequence of the odd number of derivatives. 

The metric dependence of the above expression comes from three different places. The first, which will not contain external momenta when we perform the integral in momentum space, is through the contraction of the $\alpha$ indexes between $F$ and $\tilde{F}$. The second contribution can be obtained by expanding the covariant derivatives acting on the $\mu_j$ indexes in terms of the Christoffel symbols. Those terms with derivatives acting on the Christoffel symbols will involve higher orders of the external momenta when we integrate and thus can be dropped. The remaining ones will be of the form 
\begin{equation}
\Gamma_{(\mu_i \mu_j}^\gamma \partial_{\mu_2}...\partial_{\gamma} ... \partial_{\mu_k} {F_{\mu_1}}^\alpha \partial_{\mu_{k+1}}...\partial_{\mu_{s-1-k}} \tilde{F}_{\mu_s) \alpha} \, ,  
\end{equation} 
and
\begin{equation}
\Gamma_{(\mu_i \mu_1}^\gamma \partial_{\mu_2}... \partial_{\mu_k} {F_{\gamma}}^\alpha \partial_{\mu_{k+1}}...\partial_{\mu_{s-1-k}} \tilde{F}_{\mu_s) \alpha} \, ,
\end{equation}
and the same with $F \leftrightarrow \tilde{F}$. These terms come into various combinations in the complete sum but for our purposes it is enough to argue that they will cancel term by term. To see this, one takes the functional derivative with respect to the external metric and goes to Fourier space. After performing Wick contractions, what remains is an integral of the form
\begin{equation}
\frac{1}{\beta}\sum_n \int \frac{d^3 \vec{q}}{(2\pi)^3}\left( ... \right) {\epsilon_{\mu_i}}^{\alpha\beta\gamma} L_{\mu_j \alpha \beta \gamma} (q) \, ,
\end{equation} 
where the dots stand for a combination of momenta and the rightmost part comes from the Wick contraction.
The point is that such formula vanishes identically since
\begin{equation}
{\epsilon_{\mu}}^{\alpha\beta\gamma} L_{\nu \alpha \beta \gamma}= \frac{1}{q^2} {\epsilon_{\mu}}^{\alpha\beta\gamma} \left( q_\nu q_\beta \delta_{\alpha \gamma} - q_\alpha q_\beta \delta_{\nu \gamma} - q_\nu q_\gamma \delta_{\alpha\beta} + q_{\alpha} q_{\gamma} \delta_{\nu \beta}  \right)=0.
\end{equation}
Finally, the last source of contact terms are those cases in which one acts with the covariant derivative on the contracted index $\alpha$. In position space, they give a contribution
\begin{equation}
\sum_{k=0}^{s-2} c_{s,k} \sum_{i=0}^k \Gamma_{\mu_i \alpha}^\gamma \left( \partial_{(\mu_2}... ...\partial_{\mu_{k}} {F_{\mu_1 \gamma}} \partial_{\mu_{k+1}}...\partial_{\mu_{s-1-k}} {\tilde{F}_{\mu_s)} {}^\alpha -\partial_{\mu_{k+1}}...\partial_{\mu_{s-1-k}}  {F_{\mu_1}}^\alpha \partial_{\mu_2}}...\partial_{\mu_{k}} \tilde{F}_{\mu_s) \gamma}  \right) + {\cal O} ( \partial \Gamma ) \, ,
\end{equation}
where the $\mu_{i}$-th derivative is missing. Since the indexes are all symmetrized, the sum over $i$ just gives a factor of $k$. Manipulating the binomial coefficient, one can recast the whole expression as
\begin{equation}
-\frac{s-2}{2} \Gamma_{(\mu_1 \alpha}^\gamma \left( F_{\mu_2 \gamma} \overset{\leftrightarrow}{\partial}_{\mu_3} ... \overset{\leftrightarrow}{\partial}_{\mu_{s-1}} \tilde{F}_{\mu_s)} {}^\alpha - {F_{\mu_2}}^\alpha \overset{\leftrightarrow}{\partial}_{\mu_3} ...\overset{\leftrightarrow}{\partial}_{\mu_{s-1}} \tilde{F}_{\mu_s)\gamma}  \right) + {\cal O} ( \partial \Gamma ) \, .
\end{equation}
Finally using
\begin{equation}
\frac{\delta \Gamma_{\mu\nu}^{\gamma}(x)}{\delta g_{\alpha\beta}(y)}|_{g=\delta}= \frac{1}{2} \left[ -\delta^{(\alpha}_\mu \delta^{\beta)}_\nu \partial^\gamma \delta(x-y) + \delta_\nu^{(\alpha} \delta^{\beta) \gamma}\partial_\mu \delta(x-y) + \delta_\mu^{(\alpha}\delta^{\beta)\gamma}\partial_\nu \delta(x-y) \right] \, ,
\end{equation}
one arrives at the momentum space expression for the contact terms.
\end{document}